\documentclass[10pt,twocolumn]{article}

\usepackage{latexsym}
\usepackage{algorithm}
\usepackage{algorithmic}
\usepackage{graphicx}
\usepackage{subfigure}
\usepackage[T1]{fontenc}
\usepackage{mathptmx}
\usepackage{charter}
\usepackage{amsmath}
\usepackage{amsthm}
\usepackage{fullpage}
\usepackage{complexity}
\usepackage{multirow}

\newtheorem{theorem}{Theorem}

\newcommand{\mbff}{\mathbf{f}}
\newcommand{\trm}[1]{\textrm{#1}}

\title{Improving rewards in overloaded real-time systems}

\author{
  \begin{tabular}{ccc}
    Sathish Gopalakrishnan \\
    Department of Electrical and Computer Engineering \\
    The University of British Columbia 
  \end{tabular}
}

\date{}

\begin{document}

\maketitle

\setlength{\baselineskip}{0.92\baselineskip}
\setlength{\parskip}{0.05in}
\setlength{\parindent}{0.10in}


\begin{abstract}

Competitive analysis of online algorithms has commonly been applied to understand the behaviour of real-time systems during overload conditions. While competitive analysis provides insight into the behaviour of certain algorithms, it is hard to make inferences about the performance of those algorithms in practice. Other approaches to dealing with overload resort to heuristics that seem to perform well but are hard to prove as being good. Further, most work on handling overload in real-time systems does not consider using information regarding the distribution of arrival rates of jobs and execution times to make scheduling decisions. We present an scheduling policy (obtained through stochastic approximation, and using information about the workload) to handle overload in real-time systems and improve the revenue earned when each successful job completion results in revenue accrual. We prove that the policy we outline does lead to increased revenue when compared to a class of scheduling policies that make static resource allocations to different service classes. We also use empirical evidence to underscore the fact that this policy performs better than a variety of other scheduling policies. The ideas presented can be applied to several soft real-time systems, specifically systems with multiple service classes.

\end{abstract}


\section{Introduction}

Large scale Internet-based operators provide a variety of services today. These services range from simple HTML content retrieval to sophisticated infrastructure services. Amazon.com, for example, offers a storage service (S3) for developing flexible data storage capabilities, a database with support for real-time queries over structured data (SimpleDB), and a computation cloud for web-scale computing (Elastic Cloud)~\cite{Am:08a}. Such services are offered at a basic support level, and at premium support levels with more stringent service level agreements. These SLAs specify the availability, reliability, and response times that customers can expect for the services provided. Further, several services are offered on a pay-for-use model rather than on the basis of long-term contracts.

Whereas most service providers size their systems to meet the normal demand and some spikes in workload, studies on Internet service workload have noted that peak-to-average ratio of workload varies from 1.6:1  to 6:1~\cite{Br:01a}. This large variation makes it exceedingly difficult for service providers to size their systems to handle all possible workload scenarios. Systems should, therefore, be designed to gracefully degrade under overload conditions.

Web services are illustrative of systems that need to handle heavy workload and respond to requests within bounded durations to adhere to SLAs with clients. These systems are a class of {\em soft real-time systems}. Requests for service can be associated with deadlines and revenue is accrued when a request is handled before its deadline. Missed deadlines are not catastrophic although they imply a loss in revenue.

In this article we study scheduling during periods of overload, and develop a scheduling policy for maximizing the revenue a service provider may accumulate. The revenue earned depends upon the requests serviced within the expected response times. During an overload, the system may choose to drop certain requests and (preferentially) provide service to requests from clients that offer better revenue (have opted for a higher quality of service). Under normal conditions we expect that the system is capable of handling all requests in a suitable manner.

The work presented here is applicable to those soft real-time systems where a service provider accrues revenue for every job successfully completed. To complete a job successfully, the system must meet the temporal requirements (deadline) for that job. Jobs that miss their deadlines do not produce any revenue in the model that we study. This model relies on the use of micro-payments, which are becoming a popular pricing design, and other pricing schemes can be approximated using micro-payments.

Specifically we study a system with $n$ client streams that require service. Each stream of requests consists of a sequence of jobs. Each job has an arrival time, a deadline relative to its arrival time, an execution time requirement, and a fixed reward for successful completion. These parameters would be part of the SLA between the client and the service provider. Although the SLA may indicate peak workload, the average workload might be much lower than the peak workload. Service providers multiplex service among many clients, and need to occasionally manage situations when the requests from clients overload the system; the duration of the overload maybe a few minutes to a few hours and a good scheduling policy will lead to optimal (or near-optimal) revenue for the service provider per unit time. 

We consider the behaviour of the scheduling policy over an infinite horizon. Note that a short overload duration (5 -- 10 minutes) is sufficiently long to motivate the use of infinite horizon policies when a system is receiving several hundred (or thousand) service requests per second, as is common for many Internet services. If a system were to respond to 100 service requests per minute, a 10-minute interval would yield 60000 jobs. We also aim to maximize the average reward earned per time step; this is closely related to maximizing the total reward obtained.

Through this work we have attempted to answer the following questions:
\begin{itemize}
	\item If we knew, a priori, probability distribution information about future workload, how do we develop a scheduling policy to improve revenues when a system is overloaded?
	\item Can we prove the effectiveness of such a policy?
	\item If the policy developed is optimal or near optimal, what can we understand about the performance of other scheduling policies developed (prior to this work) to handle overload situations?
	\item How much benefit do we derive from having some information about future job arrivals?
\end{itemize}

The scheduling policy we have derived is based on a stochastic improvement approach, and this approach is likely to be useful in a variety of other real-time scheduling problems.


\section{Related Work}

There has been extensive work on job scheduling for real-time systems focusing on hard real-time systems where each job has to meet its deadline; no rewards are associated with the successful completion of a job but missing a deadline could lead to safety hazards. The standard task model for a hard real-time task is a periodic task with a known period, known worst-case execution time, and a known deadline. Scheduling for these systems typically involve either static priority scheduling (rate/deadline monotonic priorities) or dynamic priority scheduling (earliest deadline first)~\cite{Li:00a,Bu:05a}. 

In the context of soft real-time systems, where real-time jobs can be executed with some flexibility, many techniques have been presented for maximizing a utility function subject to schedulability constraints. While Buttazzo, et al.~\cite{BuLiAb:05a} provide a detailed exposition on soft real-time systems, some approaches that are more closely related to the work described in this article involve the imprecise computation~\cite{ChLiLi:90a} and the IRIS (increased reward with increased service)~\cite{DeKuTo:96a} task models. In these models, a real-time job is split into a mandatory portion and an optional portion. The mandatory portion provides the basic (minimal) quality of service needed by a task; the mandatory portion has to be completed before the job's deadline. The optional part can be executed if the system has spare capacity, but it too must be completed before the job's deadline. The optional portion results in a reward, and the longer the optional portion can execute the greater is the reward garnered. The reward for executing the optional portion is described using a function of the extent to which the option portion is executed. Along these lines, Aydin, et al. presented techniques for optimal reward based scheduling for periodic real-time tasks~\cite{AyMeMo:01a}. Other techniques for maximizing utility (which can be considered as revenue/rewards) include the use of linear and non-linear optimization~\cite{SLS:98}, and heuristic resource allocation techniques such as QRAM~\cite{RLLS:97,RLLS:98}.

Our work is distinct from the imprecise computation model or the IRIS model because jobs in our task model do not have a mandatory or an optional portion. Further, a fixed revenue accrues with each job completion and this is unlike prior work we have highlighted where the reward is a function of the optional portion.

Overload in real-time systems has also received attention. Baruah and Haritsa described the {\tt ROBUST} scheduling policy for handling overload~\cite{BaHa:97a}. Baruah and Haritsa used the {\em effective processor utilization} as a measure of the ``goodness'' of a scheduling policy. The EPU is the fraction of time during an overload that the system executes tasks that complete by their deadlines. When the EPU is used as a metric for measuring the performance of a scheduling policy the task model is a {\em special case} of scheduling to improve rewards: in this model the reward for a job completion is equal to the execution time of the job. The task model studied by Baruah and Haritsa made no assumptions about the arrival rates of jobs. Each job was characterized by its arrival times, its execution time and its deadline. The {\tt ROBUST} scheduler is an optimal online scheduler among schedulers with no knowledge of future arrivals. Baruah, et al. established that no online scheduler is guaranteed to achieve an EPU greater than 0.25~\cite{BaKoMi:91a}. When the value of a job need not be related to the execution length, Baruah, et al.~\cite{BaKoMa:92a} provided a general result that the competitive ratio for an online scheduling policy cannot be guaranteed to be better than $\frac{1}{(\sqrt{k}+1)^{2}}$ where $k$ is the ratio of the largest to smallest {\em value density} among jobs to be scheduled. The value density of a job is its value-to-execution length ratio.

For systems where a job's value need not be directly related to its execution length, Koren and Shasha developed the $D^{over}$ online scheduling policy~\cite{KoSh:95a}, which provides the best possible competitive ratio relative to an offline (or clairvoyant) scheduling policy. Koren and Shasha also developed the Skip$^{over}$ scheduling approach~\cite{KoSh:95b} to deal with task sets where certain jobs can be skipped to ensure schedulability at the cost of lower quality of service. While Skip$^{over}$ was developed as a mechanism for dealing with overload, it is not suited to the application scenarios we have described earlier.

Hajek studied another special case when all jobs are unit length and concluded that the competitive ratio for online scheduling of such jobs lies in the interval $[0.5,\phi]$ where $\phi = \frac{\sqrt{5}-1}{2} \approx 0.618$, the inverse of the golden ratio~\cite{Ha:01a}.

Competitive analysis of scheduling policies provides us good insight into the behaviour of different policies but does not address all issues. The job arrival pattern that leads to poor performance of a policy $\zeta$ may be extremely rare in real systems. Additionally, two online algorithms with the same competitive ratio might have significantly varied performance in practice. Koutsoupias and Papadimitriou discuss the limitations of competitive analysis and suggest some refinements that could make problem formulation more realistic~\cite{KoPa:00a}. The limitations of competitive analysis have spurred investigations into several heuristics that offer good performance in most settings. For example, Buttazzo, et al. have described experiences with robust versions of the earliest deadline first algorithm~\cite{BuSt:93a,BuSpSe:95a}. 

With regard to prior work on handling overload in real-time systems, we study a general revenue model where the revenue earned on completing a job need not be related to the execution time of the job. Moreover, we propose a scheduling policy that has limited awareness of the characteristics of the workload. While in prior work (\cite{BaHa:97a,BaKoMi:91a,KoSh:95a,BuSt:93a,BuSpSe:95a}) no assumptions were made about future job arrivals, we use estimates of arrival rates to make better decisions. Such information can easily be measured, or specified, in a system, and is often described in the service level agreements between service providers and customers. This information is, therefore, not unreasonable to expect for the class of systems that we are interested in. Furthermore, Stankovic, et al.~\cite{StSpNa:95a} have stressed the need to incorporate more information about the workload. Writing about competitive analysis for overload scheduling (\cite{StSpNa:95a}, p. 17) they note that ``More work is needed to derive other bounds based on more knowledge of the task set.'' Although our work does not lead to deriving bounds on competitive performance of online scheduling policies, we use information concerning the task streams to develop a scheduling policy to improve revenues in the presence of overload.

Lam, et al.~\cite{LaNgTo:04a} have presented a scheme that uses faster processors to handle overload. We have proposed a scheme that is suited to situations where extra resources may not easily be available, or cannot be deployed quickly, to ameliorate overload.

Finally, we note that we use stochastic models for soft real-time systems. Real-time queueing theory~\cite{Lehoczky:96} deals with probabilistic guarantees for real-time systems but RTQT does not provide tools either for analyzing overload conditions or for maximizing rewards in a real-time system.


\section{System and task model}

The system and task model that we consider is that of $n$ streams, $\{S_{1},\dots,S_{n}\}$, with {\em preemptible} jobs; all jobs are executed on a {\em uniprocessor} system. Within a particular stream $S_{k}$ jobs arrive with a mean inter-arrival time $P_{k}$; the inter-arrival times are governed by a Poisson process with rate $r_{k} = \frac{1}{P_{k}}$.\footnote{The inter-arrival times correspond to peak workload.} The execution time of each job may also vary; for stream $S_{k}$ we consider the execution time of jobs to be governed by an exponential distribution with mean $e_{k}$. Each job also has a deadline; the deadlines for jobs of $S_{k}$ follow an exponential distribution with mean $D_{k}$. When a job belonging to $S_{k}$ is completed prior to its deadline expiring a fixed revenue of $v_{k} (> 0)$ is earned. We will use the terms revenue, value and reward interchangeably for the rest of this article.

In this work, we provide a method for achieving high average revenue over an infinite time horizon. An optimal scheduling policy, $\zeta^{*}$, is one that will achieve the supremum
\[ V^{\zeta^{*}} = \limsup_{t \rightarrow \infty} \{ \frac{V^{\zeta}(t)}{t+1} \} \] where $V^{\zeta}(t)$ is the revenue obtained using policy $\zeta$ over the interval $[0, t)$.

The scheduling policies of interest are {\em non-idling}, or work conserving, policies that make decisions whenever the state of the system changes: when a new job arrives, when a job finishes, or when a deadline expires.

This model also generalizes the traditional periodic task model studied by Liu and Layland. No relationship need exist between the deadlines and the rates of the tasks. When tasks have deterministic parameters (execution times, deadlines and periods) then the problem of dealing with an overload can be reduced to the problem of picking the subset of tasks that attains maximum revenue while eliminating the overload.

\section{Identifying a good scheduling policy}

Before we develop some intuition regarding scheduling policies that optimize the average revenue earned over a long run of the system, we note that this discussion is particularly relevant for overloaded systems, i.e., for systems where \( \sum_{i=1}^{n} \frac{e_{k}}{P_{k}} = \sum_{i=1}^{n} e_{k}r_{k} > 1. \) If the system was under-utilized then such a policy is optimal and would generate an average revenue of \( \sum_{i=1}^{n} v_{k}r_{k} \); the earliest deadline first policy, in fact, emulates this allocation when the utilization is $\le 1$. 

Whenever the system is not overloaded, we will assume the use of the EDF policy. Notice that a system is guaranteed to meet all deadlines when $\sum_{i} e_{i}/D_{i} \le 1$.

We shall identify an ideal policy by first determining an optimal static allocation of the processor among the different job streams, and then improving that allocation at each decision step. Our first goal is to determine fractional allocations of the processor among the $n$ streams. Essentially we seek a vector $\mbff = \{f_{1},f_{2},\dots,f_{n}\}$ such that $f_{i}$ represents the proportion of processor time allocated to stream $S_{i}$. In other words, such a static allocation would allocate an $f_{i}$ fraction of each time unit to task stream $S_{i}$. Although this may be an impractical policy -- because of the excessive context switching overhead -- we shall use this as an initial step to obtaining a more practical policy.

\subsection{Optimal fractional resource allocation}

We would like to partition the processor's efforts among the $n$ streams to optimize the revenue earned. $f_{i}$ represents that long-run fraction of time spent by the processor servicing jobs of stream $S_{i}$.

When dealing with systems subject to overload, job queue lengths may grow rapidly but the system is kept stable by the fact that jobs have deadlines. We let $L_{i}(t)$ represent the length of the queue of jobs from $S_{i}$ at time instant $t$. The $n$ queue lengths are stochastic processes that evolve depending on the scheduling policy chosen; further the queue lengths are independent of each other because each queue is guaranteed a fraction of the processor. The queue length $L_{i}(t)$ is, therefore, a simple birth-death process with the rate of arrivals to the queue being $r_{k}$ and the departure rate being $\frac{f_{i}}{e_{i}}+\frac{l}{D_{i}}$ [influenced by job completions and deadline expirations] when the state of the queue, the queue length, is $l$. If we use terms that are more common to queueing systems, then the service rate $s_{i} = \frac{1}{e_{i}}$, the deadline miss rate $d_{i} = \frac{1}{D_{i}}$, and the departure rate for the queue length process is $f_{i}s_{i}+ld_{i}$.

Applying some standard results concerning birth-death processes~\cite{PaPi:02a}, the stationary distribution for $L_{i}(t)$, when stream $S_{i}$ is alloted an $f_{i}$ proportion of the processor, is given by 
\begin{eqnarray}
	\Pi_{i}(l,f_{i}) & = & \frac{(r_{i})^{l}}{ \prod_{m=1}^{l} (s_{i}f_{i}+md_{i}) } \Pi_{0}(r_{i},s_{i}f_{i},d_{i}), 
\end{eqnarray} where $l$ is the state of queue $i$ and 
\begin{eqnarray}
	\Pi_{0}(r_{i},s_{i}f_{i},d_{i}) & = & \left( \sum_{l=0}^{\infty} \frac{(r_{i})^{l}}{ \prod_{m=1}^{l} (s_{i}f_{i}+md_{i}) } \right)^{-1}.
\end{eqnarray}

The average revenue obtained using scheduling policy $\zeta_{\mathbf{f}}$ that allocates $f_{i}$ proportion of the processor to stream $S_{i}$ is
\begin{eqnarray}
	V_{\mbff} & = & \sum_{i=1}^{n}v_{i}s_{i}f_{i}\left[ 1-\Pi_{0}(r_{i},s_{i}f_{i},d_{i}) \right] \label{eq:avgRevenue} 
\end{eqnarray} and the optimal fractional allocation policy $\zeta^{*}$ is that policy that picks the maximizing vector $\mathbf{f}$:
\begin{eqnarray}
	V^{*} & = & \max \{ V_{\mathbf{f}} | f_{i} \ge 0, \sum_{i=1}^{n} f_{i} = 1 \}.
\end{eqnarray}

We will initially assume that we have obtained the optimal fractional allocation policy and suggest a mechanism to improve on policies that pre-allocate processor shares. We will refer to $\zeta^{*}$, the optimal fractional allocation policy, as $FAP$. Further, we noted earlier that the fractional allocation policy might require each time step to divided among all queues, which might lead to unacceptable overhead. The improvement step will result in a policy that can be applied at every time instant when the state of the system changes, i.e., whenever a new job arrives, or when a job is completed, or when a job misses its deadline.


\subsection{An improved policy for online job selection}

We will improve upon a fractional allocation policy $\zeta_{\mathbf{f}}$ by defining a priority index $Z_{i}(l_{i})$ that indicates the priority of a stream when there are $l_{i}$ queued jobs belonging to that stream. Then, at any time $t$ when the scheduler needs to make a decision, the scheduler will activate a job from the stream with the highest priority index; thus stream $S_{j}$ will be chosen iff
\begin{eqnarray}
	Z_{j}(l_{j}) & = & \max_{i}\left\{Z_{i}(l_{i})\right\}.
\end{eqnarray}

A scheduling decision is made whenever the state of any of the queues changes. The approach underlying our improved policy is to assume that at every decision instant a particular job is scheduled and that from the next decision instant policy $FAP$ will be applied; the selection of the job at the first decision instant is based on improving the revenue in comparison to a consistent use of $FAP$. By applying the improvement step (as dictated by the priority indices) at each decision instant we can obtain consistently better performance than $FAP$. This approach can be re-stated as follows:
\begin{itemize}
	\item If $t=0$ is the first decision instant then we will select a job and execute it till the second decision instant.
	\item Assume that $FAP$ will be used from the second decision instant. Therefore, pick a job at $t=0$ that will lead to an improved revenue when compared with the use of $FAP$ from $t=0$.
	\item If we treat every decision instant exactly like the first decision instant then the modified policy will consistently outperform $FAP$.
\end{itemize}

In this article we shall denote the policy that uses the above priority index as Policy $Z$.\footnote{The name for this scheduling policy is inspired by an operating system~\cite{PiPrDo:95a} and the motion picture that influenced the operating system~\cite{Plan9}.} We shall now state the main theorem and then proceed to prove this theorem.

\begin{theorem}

The scheduling policy that improves upon the fractional allocation policy $\zeta_{\mbff}$ is the policy that chooses to service task stream $i$ when $l_{i} > 0$ and 
\[
	Z_{i}(l_{i}) = \max_{j; l_{j} > 0}\{Z_{j}(l_{j})\}
\] where 
\[
	Z_{i}(l_{i}) = v_{i}s_{i}\left[1-\frac{(s_{i}f_{i})\Pi_{0}(r_{i},s_{i}f_{i},d_{i})}{(s_{i}f_{i}+l_{i}d_{i})\Pi_{0}(r_{i},s_{i}f_{i}+l_{i}d_{i},d_{i})}\right].
\]

\label{th-main}
\end{theorem}

{\em Understanding the modified policy.} The prioritization suggested by the updated scheduling policy is greedy. This is expected when scheduling tasks with deadlines. The priorities are based on the highest possible revenue rate ($v_{i}s_{i}$). At the same times, the priority attempts to delay those streams that typically have longer deadlines; draining queues that have jobs that can wait would, at later time instant, lead to serving jobs that do not yield high revenues and this is reflected by the zero probability term $\Pi_{0}(r_{i},s_{i}f_{i},d_{i})$. However, if a queue is sufficiently long then we can serve jobs in that queue without worrying about draining that queue and this is reflected by the $\Pi_{0}(r_{i},s_{i}f_{i}+l_{i}d_{i},d_{i})$ term. Also, when deadlines are short the deadline miss rate ($d_{i}$) is high and this is captured by the term $(s_{i}f_{i}+l_{i}d_{i})^{-1}$ that boosts the priority of streams with shorter deadlines.

Whenever a scheduling decision is to be made, the optimal choice would depend on whether executing a job now is better than deferring its execution. The penalty that one may incur by deferring the execution of a job is that the job may miss its deadline thereby resulting in no revenue. We denote the expectation of the revenue earned from $S_{i}$ by applying the fractional allocation policy when the state of queue $i$ is $l$ as $V_{\mbff,i}(l)$. The priority of each stream can then be computed as
\begin{eqnarray}
	Z_{i}(l_{i}) & = & s_{i}[v_{i} - \{V_{\mbff,i}(l_{i})-V_{\mbff,i}(l_{i}-1)\}].
	\label{eq-Main}
\end{eqnarray}

{\em Proof outline.} In computing the priorities we essentially account for the potential loss in revenue if we defer the execution of a job to a later time instant. The highest priority job is that job that will result in the maximum loss if its execution were to be deferred and its deadline were to expire as a consequence of the deferral. It becomes essential to compute the expected change in revenue, $V_{\mbff,i}(l_{i})-V_{\mbff,i}(l_{i}-1)$ before we can determine the priority of a job. The rest of this section is dedicated to a discussion on how we can recover this quantity.

To understand the long-run average reward obtained from a particular class of workload, we consider the evolution of the queue $\{L_{i}(t), t \in R^{+}\}$ with initial condition $L_{i}(0) = l_{i}$ and being awarded a fraction $f_{i}$ of processing time. The queue length will evolve as a birth-death process with birth rate $r_{i}$ and death rate $s_{i}f_{i}+ld_{i}$ at time $t$ with $l \in Z^{+}, l = L_{i}(t)$.

A scheduling policy that apportions fractional processing to different job streams is guaranteed an average revenue of $f_{i}s_{i}v_{i}$ from stream $i$ as long as queue $i$ is never empty. If we have determined the optimal fractional allocations then a scheduling policy can attain high value by not allowing queues to empty: jobs that provide high revenue and have short deadlines may be preferred. We will, therefore, understand the variation in the emptying time of a queue if a job is processed at time instant $t$ or at a later time instant.

The remainder of the proof is devoted to identifying the quantity $V_{\mbff,i}(l_{i})-V_{\mbff,i}(l_{i}-1)$.

\begin{proof}

The stopping time for the birth-death process $\{L_{i}(t),t \in R^{+}\}$ when the scheduling policy uses fractional allocations defined by the vector $\mbff$ is defined as 
\begin{eqnarray}
	\tau_{\mbff,i}(l_{i}) & := & \inf\{ t | t > 0 \trm{ and } L_{i}(t) = 0 \}.
\end{eqnarray}
The expected value obtained from queue $i$ in the interval $[0,\tau_{\mbff,i}(l_{i}))$ is denoted $\hat{V}_{\mbff,i}(l_{i})$. Further, we denote the expectation for the stopping time as
\begin{eqnarray}
	\overline{T}_{\mbff,i}(l_{i}) & := & E[\tau_{\mbff,i}(l_{i})].
\end{eqnarray}

From standard results concerning Markov Decision Processes~\cite{Pu:05a}, we can establish that the $0$ state is a regeneration point for the queuing process $\{L_{i}(t), t \in R^{+}\}$. We can then obtain 
\begin{equation}
	\begin{tabular}{l}
		$V_{\mbff,i}(l_{i})-V_{\mbff,i}(l_{i}-1) = $\\
		~~~~~~ $\hat{V}_{\mbff,i}(l_{i})-\hat{V}_{\mbff,i}(l_{i}-1)-[\overline{T}_{\mbff,i}(l_{i})-\overline{T}_{\mbff,i}(l_{i}-1)] V_{\mbff,i},$ \\ 
		~~~~~~ $l_{i} \in Z^{+}, ~~~ 1 \le i \le N.$
	\end{tabular}
	\label{eq-valueDiff}
\end{equation}

Notice that if we define an alternative stopping time
\begin{eqnarray}
	\hat{\tau}_{\mbff,i}(l_{i}) & := & \inf\{t | t > 0 \trm{ and } L_{i}(t) = l_{i}-1\}
\end{eqnarray}
then \( \hat{V}_{\mbff,i}(l_{i})-\hat{V}_{\mbff,i}(l_{i}-1) \) is the value derived from servicing queue $i$, which is governed by the MDP $\{L_{i}(t), t \in R^{+}\}$ with $L_{i}(0) = l_{i}$ during the interval $[0, \hat{\tau}_{\mbff,i}(l_{i}))$. 

Also,
\begin{eqnarray}
\overline{T}_{\mbff,i}(l_{i})-\overline{T}_{\mbff,i}(l_{i}-1) & = & E[\hat{\tau}_{\mbff,i}(l_{i})].
\end{eqnarray}

We shall now introduce a shadow process $\{\hat{L}_{i}(t), t \in R^{+}\}$ to ease our analysis. This process {\em shadows} the queueing process $\{L_{i}(t), t \in R^{+}\}$ with some subtle differences. The shadow process is a birth-death process with birth rate $r_{i}$ and death rate $s_{i}f_{i}+(l_{i}+m)d_{i}$ in state $(l_{i}+m), m \in N$. The death rate is $0$ in states where the queue length is less than $l_{i}$. The initial state of the shadow process is $\hat{L}_{i}(0) = l_{i}$. The shadow process is identical to the original queue length process $\{L_{i}(t), t \in R^{+}\}$ when the queue length is greater than $l_{i}-1$ but the shadow process cannot enter the state where the queue length is $l_{i}-2$. The shadow process has as its regeneration point the state $l_{i}-1$ and the reward derived from the shadow process per unit time is
\begin{eqnarray}
	\tilde{V}_{\mbff,i} & = & \frac{\hat{V}_{\mbff,i}(l_{i})-\hat{V}_{\mbff,i}(l_{i}-1)}{r_{i}^{-1}+\overline{T}_{\mbff,i}(l_{i})-\overline{T}_{\mbff,i}(l_{i}-1)}.
	\label{eq-value1}
\end{eqnarray}

In the expression for $\tilde{V}_{\mbff,i}$, the numerator represents the reward earned when the original MDP transitions from state $l_{i}$ to $l_{i-1}$; the denominator is the expected duration for the shadow process to return to its initial state, i.e., start from the initial state of $l_{i}$, transition to state $l_{i-1}$ and then return to state $l_{i}$.

From standard results regarding birth-death processes~\cite{PaPi:02a} we can obtain the stationary distribution for $\{\hat{L}_{\mbff,i}(t), t \in R^{+}\}$ as 
\begin{eqnarray}
	\hat{\Pi}_{i}(l) & = & \left\{ 
			\begin{array}{ll}
				\frac{\Pi_{0}(r_{i},s_{i}f_{i}+(l_{i}-1)d_{i},d_{i})}{\prod_{m=l_{i}}^{l} (s_{i}f_{i}+md_{i})}r_{i}^{l-l_{i}+1}, & l \ge l_{i}-1 \\
				0, & l \le l_{i}-2
			\end{array}
		\right.
\end{eqnarray}

The value obtained per unit time for the shadow process, which does not earn any revenue in state $l_{i}-1$, is given by
\begin{equation}
	\begin{tabular}{l}
	$v_{i}s_{i}f_{i}(1-\hat{\Pi}_{i}(l_{i}-1)) =$
	\\ ~~~~~~ $v_{i}s_{i}f_{i}[1-\Pi_{0}(r_{i},s_{i}f_{i}+(l_{i}-1)d_{i},d_{i})].$
	\end{tabular}
	\label{eq-A}
\end{equation}

Further, we can use \eqref{eq:avgRevenue} and \eqref{eq-value1} to infer that
\begin{equation}
	\begin{tabular}{l}
	$\frac{(\hat{V}_{\mbff,i}(l_{i})-\hat{V}_{\mbff,i}(l_{i}-1))}{r_{i}^{-1}+\overline{T}_{\mbff,i}(l_{i})-\overline{T}_{\mbff,i}(l_{i}-1)} =$ \\
	~~~~~~ $v_{i}s_{i}f_{i}[1-\Pi_{0}(r_{i},s_{i}f_{i}+(l_{i}-1)d_{i},d_{i})]$
	\end{tabular}
	\label{eq-B}
\end{equation}
and that 
\begin{equation}
	\begin{tabular}{l}
	$\frac{(\overline{T}_{\mbff,i}(l_{i})-\overline{T}_{\mbff,i}(l_{i}-1))}{r_{i}^{-1}+\overline{T}_{\mbff,i}(l_{i})-\overline{T}_{\mbff,i}(l_{i}-1)} =$ \\
	~~~~~~ $1-\Pi_{0}(r_{i},s_{i}f_{i}+(l_{i}-1)d_{i},d_{i}).$
	\end{tabular}
	\label{eq-C}
\end{equation}

We can now combine \eqref{eq-valueDiff}, \eqref{eq-A}, \eqref{eq-B} and \eqref{eq-C} to conclude that 
\begin{equation}
	\begin{tabular}{l}
	$V_{\mbff,i}(l_{i})-V_{\mbff,i}(l_{i}-1) =$ \\
	~~~~~~ $\frac{[v_{i}s_{i}f_{i}][\Pi_{0}(r_{i},s_{i}f_{i},d_{i})]}{[s_{i}f_{i}+l_{i}d_{i}][\Pi_{0}(r_{i},s_{i}f_{i}+l_{i}d_{i},d_{i})]}.$
	\end{tabular}
	\label{eq-Main2}
\end{equation}

Finally, we use \eqref{eq-Main2} in \eqref{eq-Main} to complete the theorem.

\end{proof}


\section{Empirical evaluation}

Having described the structure of a policy for job selection to maximize rewards, we shall now describe simulation results that compare the performance of our policy with other approaches.  

Before elaborating on empirical evaluation, we emphasize that it is extremely difficult to exhaustively evaluate, via simulation, different scheduling policies, especially when rewards can be assigned arbitrarily. The proof that Policy $Z$ can yield strong, and increased, revenue (Theorem~\ref{th-main}) is what should suggest the ``goodness'' of the policy. The empirical evaluations are only indicative of the general applicability of that result.

\subsection{Comparison with stochastic dynamic programming}

Optimal solutions to the scheduling problem of interest can be recovered using stochastic dynamic programming~\cite{Ro:95a}. Stochastic dynamic programming is, however, computationally expensive and is not practical for most applications. For a simple workload with at most {\em two} task streams it is computationally feasible to resort to SDP; we used this case to compare the performance of the proposed policy with the optimal policy.

We begin by making two comparisons: 
\begin{enumerate}
	\item Optimal fractional allocation ($FAP$) vs. Policy $Z$, and
	\item Policy $Z$ vs. the optimal policy via SDP.
\end{enumerate}

For these comparisons we used many task streams, and we present the results from a representative set of simulation runs (parameters in Table~\ref{t:ZvsOPT}). Each run consisted of two task streams, and the simulations were performed for $9,00,000$ time steps. Each task stream had the same average inter-arrival time of 350 time units, and the revenue earned for every job of task stream 2 was 1.0, i.e., $v_{2}=1.0$. We also kept the same mean deadline for each task stream, $D=D_{1}=D_{2}$. For some simulation runs the mean execution time was longer than the mean deadline, making scheduling decisions even harder.

\begin{table*}[htb]
\begin{center}
	\begin{tabular}{|cccccc|} \hline
		Experiment & $D$ & $e_{1}$ & $e_{2}$ & $v_{1}$ & $f_{1}^{*}$ \\ \hline \hline
		$E_{1}$ & 1000 & 600 & 600 & 1.0 & 0.50 \\
		$E_{2}$ & 1000 & 620 & 725 & 1.1 & 0.54 \\
		$E_{3}$ & 1000 & 580 & 790 & 1.2 & 0.57 \\
		$E_{4}$ & 1000 & 545 & 855 & 1.3 & 0.61 \\
		$E_{5}$ & 1000 & 520 & 925 & 1.4 & 0.64 \\
		$E_{6}$ & 1000 & 500 & 1010 & 1.5 & 0.67 \\
		$E_{7}$ & 500 & 610 & 735 & 1.1 & 0.55 \\
		$E_{8}$ & 500 & 530 & 900 & 1.3 & 0.63 \\
		$E_{9}$ & 500 & 475 & 1110 & 1.5 & 0.70 \\
		$E_{10}$ & 250 & 590 & 765 & 1.1 & 0.56 \\
		$E_{11}$ & 250 & 495 & 1020 & 1.3 & 0.67 \\
		$E_{12}$ & 250 & 435 & 1430 & 1.5 & 0.77 \\
		$E_{13}$ & 165 & 575 & 785 & 1.1 & 0.58 \\
		$E_{14}$ & 165 & 465 & 1170 & 1.3 & 0.72 \\
		$E_{15}$ & 165 & 400 & 2000 & 1.5 & 0.84 \\
		\hline
	\end{tabular}
\caption{Task stream parameters to compare the performance of the proposed policy with other policies (two task streams)}
\label{t:ZvsOPT}
\end{center}
\end{table*}

\begin{figure*}[htb]
\begin{center}
	\resizebox{14cm}{!}{
		\includegraphics{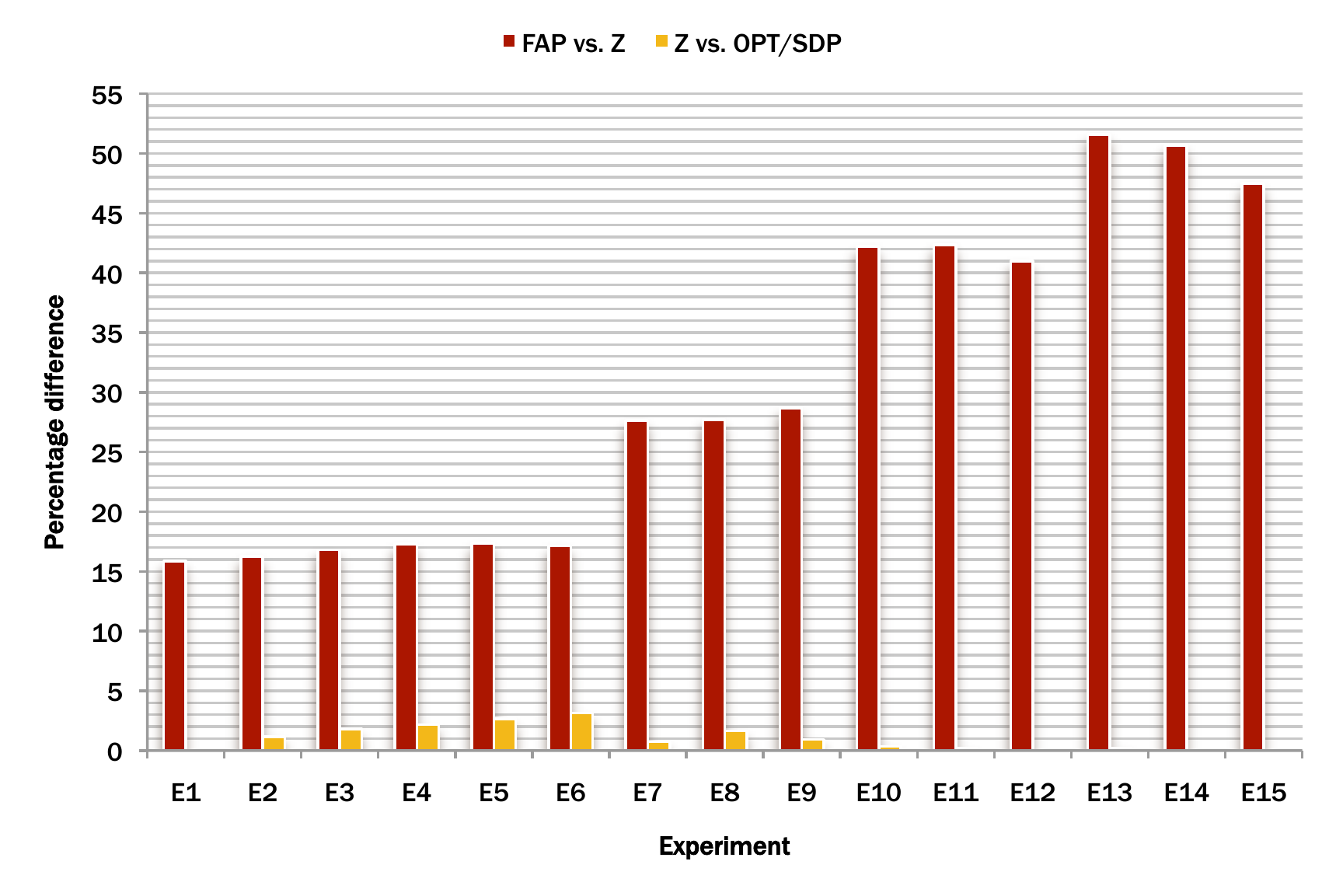}
	}
	\caption{Performance of Policy $Z$ compared with the optimal fractional policy and SDP}
	\label{f:ZvsOPT}
\end{center}
\end{figure*}

We describe our results for each experiment (Figure~\ref{f:ZvsOPT}). The optimal fractional allocation is described with other parameters (Table~\ref{t:ZvsOPT}). Recall that $f_{2}^{*}=1-f_{1}^{*}$. Policy $Z$ clearly improves over $FAP$; the {\em percentage improvement} in average revenue is at least $15\%$ (red bars in the graph). We compute the percentage improvement as follows: If $V_{Z}$ was the revenue accrued by Policy $Z$ at the end of an experiment and if $V_{FAP}$ was the reward accrued using $FAP$, then the percentage improvement is $100 \times \frac{V_{Z}-V_{FAP}}{V_{FAP}}$. 

In comparison to the optimal policy recovered using SDP, we determined the {\em loss in average revenue}  (percentage loss = $100 \times \frac{V_{SDP}-V_{Z}}{V_{SDP}}$) using policy $Z$ (yellow bars); the maximum loss was not more than $4\%$. This confirms the dramatic improvement that can be obtained over the $FAP$ and indicates that the suggested policy has a performance that is very close to the optimal SDP policy. The performance of Policy $Z$ improves when the rate of deadline misses increases.

\subsection{Comparison with {\tt ROBUST}}

Baruah and Haritsa developed the {\tt ROBUST} scheduler~\cite{BaHa:97a} for achieving near-optimal performance during overload for a specific class of systems where
\begin{itemize}
	\item The value of a job is equal to its execution length, and
	\item Each job has a slack of at least $s$, i.e., $\frac{D_{i}}{e_{i}} \ge s$.
\end{itemize}

The performance of the {\tt ROBUST} scheduler is near-optimal in the sense that it can, asymptotically, match the performance of the optimal online scheduling policy for the mentioned class of systems. They showed that the best performance that an online scheduler can guarantee is an EPU of $\lceil s \rceil/(\lceil s \rceil + 1)$ and that the {\tt ROBUST} scheduler guarantees an EPU that is at most $2/s(s+1)$ fractionally off from the optimum~\cite{BaHa:97a}.

We provide a brief description of the {\tt ROBUST} scheduler before detailing some empirical comparisons between the Policy $Z$ and {\tt ROBUST}. The {\tt ROBUST} scheduler partitions an overloaded interval into an even number of contiguous phases (Phases-$1, \dots, 2a$). The length of each even numbered phase is equal to a $1/(s-1)$ fraction of the length of the preceding odd numbered phase. At the start of an odd phase, the algorithm selects the longest eligible job and executes it non-preemptively. This job may have been executed in the previous even numbered phase; the length of the odd numbered phase is equal to the execution time remaining for that job. An odd phase concludes with the termination of the chosen job. During an even numbered phase, the scheduler selects a job with maximum length; this job may be preempted if another job arrives with longer execution length.

To compare Policy $Z$ with the {\tt ROBUST} scheduler, we used several simulations. For two sets of simulated runs, we chose a fixed slack factor of 2; for the other two sets of runs we chose a slack factor of 4. Each simulated run lasted 1,000,000 time units and involved four task streams. The execution time for jobs belonging to the same task stream were drawn from the same exponential distribution (the mean execution times for the four task streams were 50, 100, 200 and 400 respectively); the deadline for each job was set based on the slack factor. For simplicity we chose the same arrival rate for all streams; based on the desired workload intensity the arrival rate was determined.\footnote{We did perform a variety of simulation studies with different arrival rates for different task streams. To keep the article pertinent and brief, we have avoided listing all studies. The performance of the scheduling policies when the arrival rates for different streams are different is similar to the results reported in this article.} Only Policy $Z$ is concerned with task streams; the {\tt ROBUST} scheduler simply schedules on a job-by-job basis. The reward for completing a job successfully was equal to the execution time for that task stream. We do not intend this empirical analysis to be exhaustive but merely indicative of the benefits of using stochastic approximation to derive scheduling policies. For each data point, we averaged 50 independent simulation runs and compared the behaviour of the two policies.

\begin{figure*}[htb]
\begin{center}
	\resizebox{12cm}{!}{
		\includegraphics{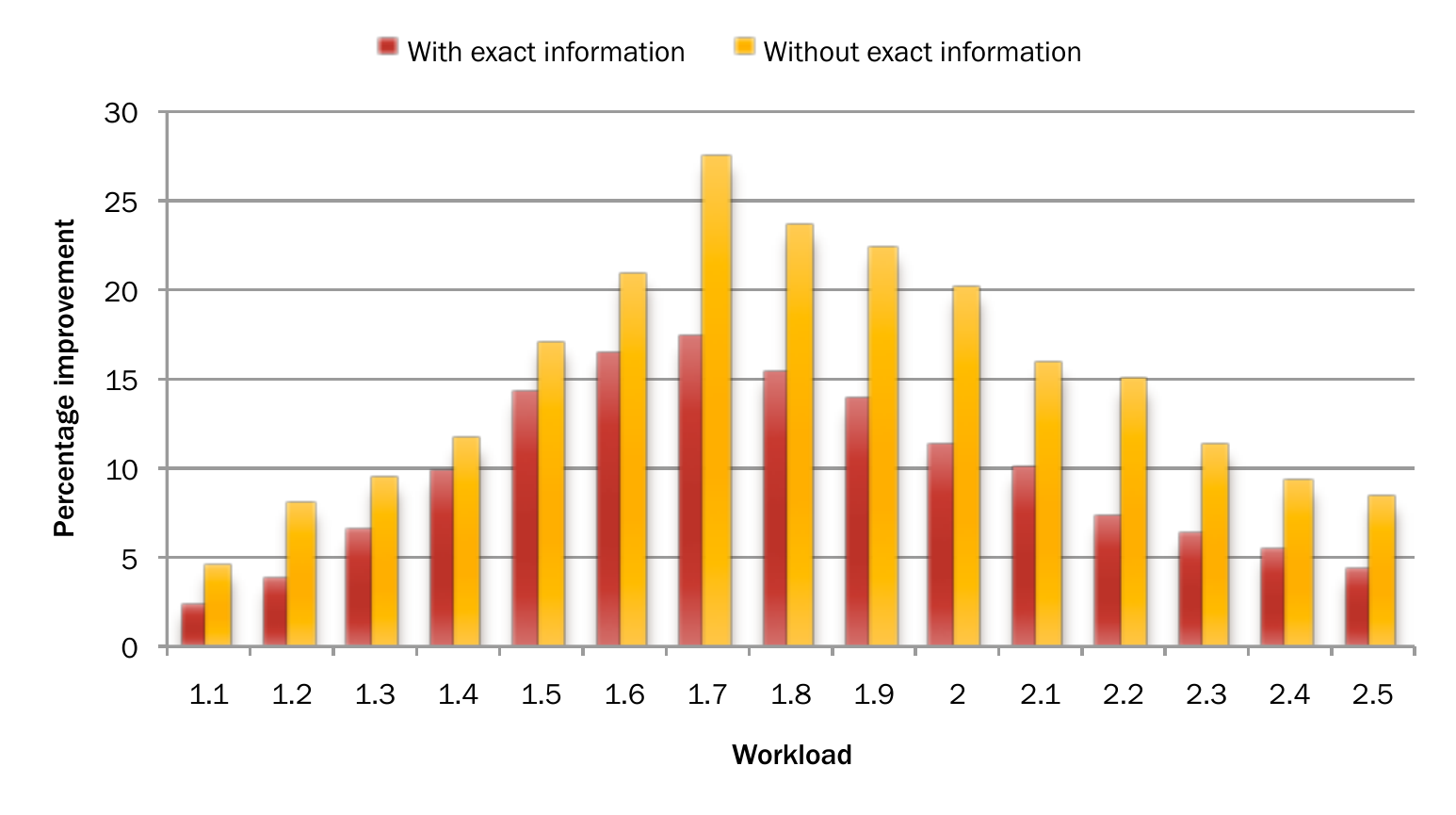}
	}
	\caption{Performance of Policy $Z$ compared with the {\tt ROBUST} policy when slack factor is 2}
	\label{f:ZvsROBUST-1}
\end{center}
\end{figure*}

\begin{figure*}[htb]
\begin{center}
	\resizebox{12cm}{!}{
		\includegraphics{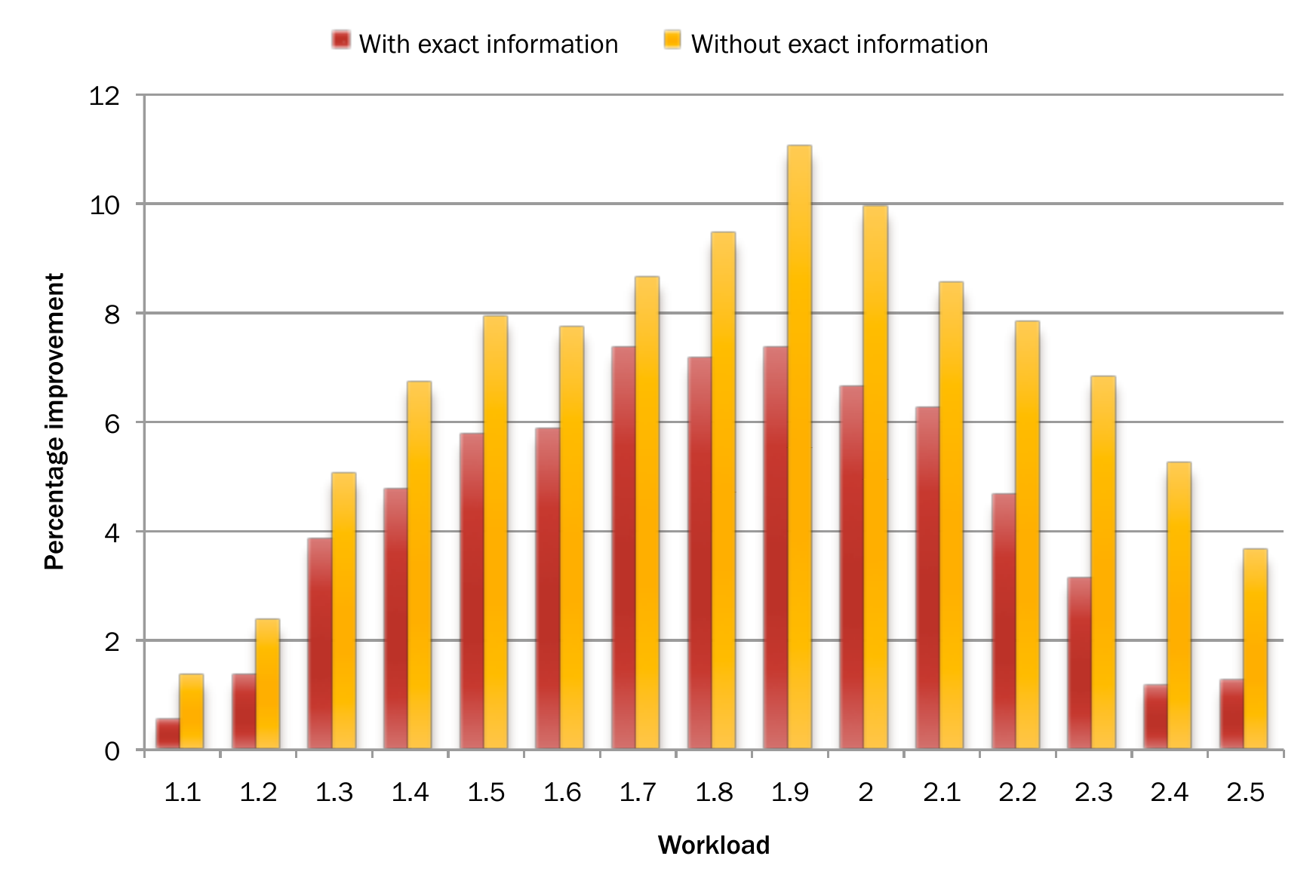}
	}
	\caption{Performance of Policy $Z$ compared with the {\tt ROBUST} policy when the slack factor is 4}
	\label{f:ZvsROBUST-2}
\end{center}
\end{figure*}

We found that Policy $Z$ outperformed {\tt ROBUST} in all scenarios (Figures~\ref{f:ZvsROBUST-1} and \ref{f:ZvsROBUST-2}). Policy $Z$ is not clairvoyant, but the awareness of potential future arrivals enables it to make better decisions. With a slack factor of 2 ($s=2$), we were able to improve the per-time step rewards in excess of $15\%$ in some cases. When the slack factor increases ($s=4$), Policy $Z$ was able improve revenue per time step but the increases are smaller. When the slack factor is high, most policies will be able to recover from a poor decision and still generate near-optimal revenue.

The {\tt ROBUST} scheduler requires accurate knowledge of the execution times of jobs and their deadlines. Policy $Z$ is obtained via stochastic approximation and is more tolerant of errors in the parameters. When the {\tt ROBUST} scheduler is only provided with the mean execution time for a job its performance drops significantly and the improvement noticed by using Policy $Z$ is more pronounced. (The red bars in Figures~\ref{f:ZvsROBUST-1} and \ref{f:ZvsROBUST-2} are based on the {\tt ROBUST} scheduler using exact information; the orange bars are based on approximate information.)

Another observation is that when the extent of overload is small, both policies perform equally well (or equally poorly). Similarly, when the system experiences heavy overload, most choices are equally good and the two policies have smaller differences.

\subsection{Comparison with {\tt REDF}}

The {\tt ROBUST} scheduler is targeted at systems with known slack factors and with a job's value being equal to its execution time. The policy we have developed, however, is also suited to arbitrary reward assignments and to situations when jobs do not have a guaranteed slack.

To understand the performance of Policy $Z$ under general workloads we compared its performance with the performance offered by the Robust EDF heuristic~\cite{BuSt:93a,BuSpSe:95a}. The {\tt REDF} policy is identical to EDF when the system is not overloaded. Whenever a job arrives a check is performed to determine if the system is overloaded. (If tasks are scheduled using EDF and $e_{i}/D_{i} \le 1$ then the system is not overloaded.) When an overload is detected, the least value task that can prevent the system from being overloaded is removed from the queue of pending jobs to a reject queue.\footnote{This policy can be modified and a smart search strategy might remove multiple jobs of low value to prevent overload. We have not implemented this approach in our evaluation.} If some job completes ahead of time then jobs from the reject queue whose deadlines have not expired may be brought back to the pending queue. Buttazzo, Spuri and Sensini showed that {\tt REDF} is well behaved during overloads~\cite{BuSpSe:95a}, and we used additional simulations to understand the performance of {\tt REDF} and Policy $Z$, and to contrast the two approaches.

\begin{figure*}[htb]
\begin{center}
	\resizebox{12cm}{!}{
		\includegraphics{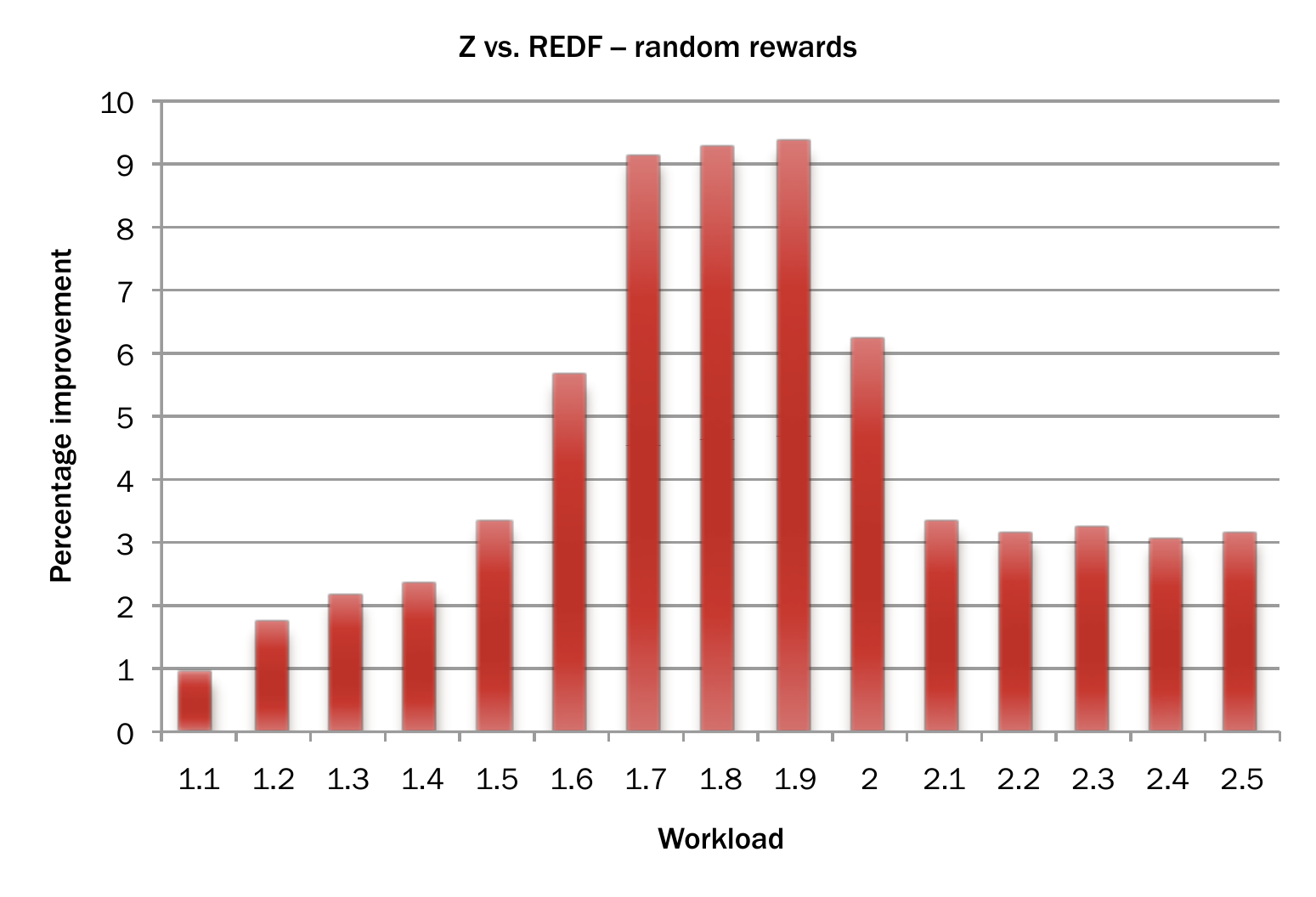}
	}
	\caption{Performance of Policy $Z$ compared with the {\tt REDF} policy (random rewards)}
	\label{f:ZvsREDF-1}
\end{center}
\end{figure*}

For these simulations, we used the task streams similar to those in our comparisons with {\tt ROBUST}. For each run we used four task streams, $S_{1}, S_{2}, S_{3}, S_{4}$, with mean execution times of 150, 100, 200 and 400 respectively. The deadlines for jobs of the four task streams were drawn from exponential distributions with mean 600, 800, 1600 and 3200 respectively. The arrival rate was chosen to generate the required workload. Similar to the previous evaluation, each stream had the same arrival rate. 

We compared the performance of {\tt REDF} with Policy $Z$ under two reward models:
\begin{itemize}
	\item The rewards associated with jobs of the four streams were 150, 300, 400 and 200 respectively. These were chosen to represent a {\em random} ranking of task streams in terms of value. 
	\item The reward associated with each stream was inversely related to the mean deadline for that stream, i.e., shorter the deadline greater the reward. The rewards associated with $S_{1}, \dots, S_{4}$ were 450, 300, 200 and 100 respectively. This reward model was intended to be approximately {\em linear} in terms of job deadlines. 
\end{itemize}

\begin{figure*}[htb]
\begin{center}
	\resizebox{12cm}{!}{
		\includegraphics{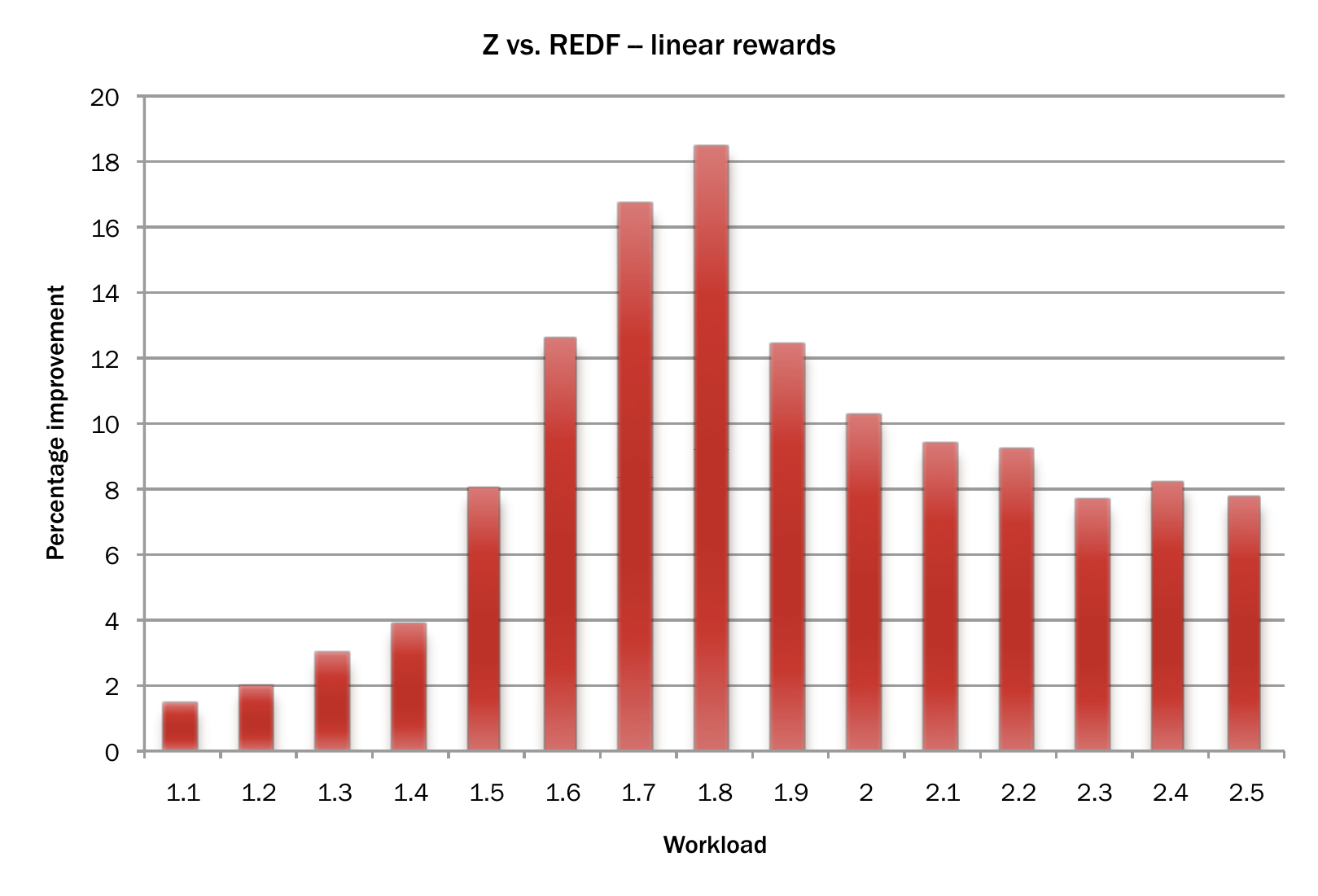}
	}
	\caption{Performance of Policy $Z$ compared with the {\tt REDF} policy (linear rewards)}
	\label{f:ZvsREDF-2}
\end{center}
\end{figure*}

We note that Policy $Z$ results in measurable improvements in revenue when compared with {\tt REDF} using both the {\em random} (Figure~\ref{f:ZvsREDF-1}) and the {\em linear} (Figure~\ref{f:ZvsREDF-2}) reward models. The linear reward model indicates greater differences because {\tt REDF} has to choose to drop jobs that may yield high rewards (because higher rewards are connected to higher utilization, and one job providing high reward may dropped in place of multiple jobs that jointly yield a smaller reward) to ensure that other jobs meet their deadlines. 

\subsection{Discussion}

On the basis of the three different comparisons (with SDP, with {\tt ROBUST}, with {\tt REDF}), we were able to ascertain the uniformly improved performance that the proposed scheduling approach (Policy $Z$) is able to offer. These comparisons strongly indicate that using knowledge of future workload does increase the revenue one can earn. The improvement in revenue can be at least $10\%$, and is likely higher when perfect information regarding the temporal requirements of jobs is not available. The improvements in revenue obtained using Policy $Z$ diminish when the system is extremely overloaded; this hints at the possibility that most scheduling decisions are likely to be reasonable in those situations.

We speculate that if Policy $Z$ is near optimal (as is the case when there are two task streams -- see Figure~\ref{f:ZvsOPT}) then other scheduling policies (e.g., {\tt ROBUST}, {\tt REDF}) are also likely to be only about 20 to 25\% away from optimality (even less in some cases) in practice, and that is an encouraging result concerning the practical applicability of those policies.

The structure of the priority index for Policy $Z$ is intuitive and can form the basis for obtaining good scheduling heuristics even when workload might not conform to simple probability distributions.

{\em Implementation considerations.} Policy $Z$ requires a priority for each class of requests, and this dynamic priority depends on the length of the corresponding queues. It is possible to compute the priorities at different queue lengths offline and use a table lookup to identify the priorities of tasks online. This makes the proposed policy easy to implement. We also need to identify the optimal fractional allocation policy, and this is also an offline operation. Identifying the optimal fractional allocation is an optimization problem in itself and we use a search over the space of possible allocations to determine the optimal allocation. This is feasible when the number of service classes is limited. It is likely that some sub-optimal initial allocations may not affect the behaviour of Policy $Z$ significantly but this notion requires further study.

\section{Conclusions}

Overload in certain soft real-time systems (such as Internet-based services) is often unavoidable because the costs of provisioning for peak load are significantly greater than the costs of handling typical load. In such systems, service providers need to provide the best possible service to customers who demand higher quality of service and are willing to pay more for better QoS. We have presented a scheduling policy for handling overload conditions and improving the revenue earned by using information about future job arrivals. The policy that we present, Policy $Z$, is based on stochastic approximation. It is not a fully clairvoyant policy and does not require accurate information about future arrivals to make scheduling decisions; approximate information about future workload is sufficient to make good decisions. 

Policy $Z$ is provably better than some policies, and empirical evidence suggests excellent performance when compared with other scheduling policies for value maximization in the presence of overload.Our policy is also sufficiently general and can be used in multiprocessor systems as well. We have restricted the discussion in this article to uniprocessor systems but it is easy to use the policy in a system with $m$ processors by selecting the top $m$ jobs based on their priority indices.

Although we make some assumptions about job arrival rates and deadlines, we believe that the approach of generating an initial policy and then improving upon that policy (as we do with the optimal fractional allocation policy and Policy $Z$) is a useful tool for decision making in real-time systems that can be generalized and applied to other problems as well.


\bibliographystyle{acm}
\setlength{\baselineskip}{1.2\baselineskip}
\bibliography{../../Bibliography/CompleteBibliography}



\end{document}